\documentclass[runningheads]{llncs}
\usepackage[T1]{fontenc}
\usepackage{float}
\usepackage{graphicx}
\usepackage{hyperref}
\usepackage{color}
\usepackage[table]{xcolor}
\usepackage{array}
\usepackage{xurl}

\usepackage{tabularx}
\usepackage{amsmath}
\begin{document}
\title{Leveraging Pathology Foundation Models for Panoptic Segmentation of Melanoma in H\&E Images}
\titlerunning{Leveraging Foundation Models for Tissue Segmentation}
\author{Jiaqi Lv\inst{1} \and
Yijie Zhu\inst{1} \and
Carmen Guadalupe Colin Tenorio\inst{3,4} \and 
Brinder Singh Chohan\inst{2} \and
Mark Eastwood\inst{1} \and
Shan E Ahmed Raza\inst{1}
}
% \author{Anonymous authors
% }
%
\authorrunning{J. Lv et al.}
% \authorrunning{Anonymous authors}
\institute{Tissue Image Analytics Centre, Department of Computer Science, University of Warwick, Coventry, United Kingdom
\email{tia@warwick.ac.uk}
\and
University Hospitals of Derby and Burton NHS Foundation Trust, United Kingdom 
\and
Department of Pathophysiology and Allergy Research, Center for Pathophysiology, Infectiology and Immunology, Medical University of Vienna, Austria
\and 
TissueGnostics GmbH, Vienna, Austria
}
% \institute{Anonymous Institutions
% }
%
\maketitle              % typeset the header of the contribution
\begin{abstract}
Melanoma is an aggressive form of skin cancer with rapid progression and high metastatic potential. Accurate characterisation of tissue morphology in melanoma is crucial for prognosis and treatment planning. However, manual segmentation of tissue regions from haematoxylin and eosin (H\&E) stained whole-slide images (WSIs) is labour-intensive and prone to inter-observer variability, this motivates the need for reliable automated tissue segmentation methods. In this study, we propose a novel deep learning network for the segmentation of five tissue classes in melanoma H\&E images. Our approach leverages Virchow2 \cite{ref_virchow2}, a pathology foundation model trained on 3.1 million histopathology images as a feature extractor. These features are fused with the original RGB images and subsequently processed by an encoder-decoder segmentation network (Efficient-UNet) to produce accurate segmentation maps. The proposed model achieved first place in the tissue segmentation task of the PUMA Grand Challenge \cite{ref_puma_paper}, demonstrating robust performance and generalizability. Our results show the potential and efficacy of incorporating pathology foundation models into segmentation networks to accelerate computational pathology workflows.

\end{abstract}
\section{Introduction}
Melanoma is an aggressive form of skin cancer with rapid progression, high metastatic potential, and poor prognosis, particularly at advanced stages \cite{ref_melanoma_staging}. Effective clinical management of melanoma relies on accurate diagnosis and precise characterization of tissue morphology from histopathology slides. Previous studies have highlighted that tumour-infiltrating lymphocytes (TILs) can potentially serve as a strong prognostic biomarker \cite{ref_taube,ref_weiss}. Specifically, the density and spatial organization of TILs can provide insights into tumour progression, response to treatment, and patient survival \cite{bruni2020tumor,hendry2017assessing,saltz2018spatial}. However, manual assessment of TILs from haematoxylin and eosin (H\&E)-stained whole-slide images (WSIs) is labour-intensive, time-consuming, and susceptible to inter-observer variability. This underscores the need for reliable automated methods for precise and reproducible tissue segmentation and cell detection in histopathological images.

Automated computational approaches using deep learning have emerged as promising solutions to address these challenges and have demonstrated strong performance in various histopathology tasks, including nuclei detection, tissue segmentation, and WSI-level classification \cite{srinidhi2021deep}. Among these methods, U-Net remains popular and continues to achieve state-of-the-art (SOTA) results in many medical image segmentation tasks, primarily due to its ability to extract multi-level features and the skip connections between its encoder and decoder \cite{ref_unet_paper}. A more recent work, nnUNet, optimizes the U-Net architecture by automatically selecting dataset-specific hyperparameters, streamlining the segmentation pipeline and achieving consistently high performance in several biomedical imaging benchmarks \cite{ref_nnunet_paper}. However, its fully automated pipeline reduces flexibility, restricting researchers from modifying the model architectures or employing tailored training strategies.

Advancements in computer vision have shown that vision transformers (ViTs) can surpass traditional convolutional neural networks (CNNs) on natural image recognition benchmarks such as ImageNet \cite{dosovitskiy2020image}. Nevertheless, ViTs are inherently data-hungry, they require large-scale annotated datasets for effective training, which poses a significant challenge in computational pathology due to the costs and labour involved in creating extensive, high-quality annotations, especially for rare cancers or tissues. For example, prior to the recently released PUMA Challenge dataset \cite{ref_puma_paper}, there were no publicly available datasets specifically for melanoma tissue segmentation, severely limiting research progress in this area.  

Pathology foundation models have recently emerged as a promising solution to the challenges of computational pathology. Foundation models are typically based on ViTs, these models are trained on millions of unlabelled or sparsely annotated histopathology images through self-supervised or weakly-supervised learning, thereby circumventing the need for extensive manual annotations. Examples of such foundation models include UNI \cite{ref_UNI_paper}, Phikon-v2 \cite{ref_phikonv2}, TITAN \cite{ref_titan}, and Virchow2 \cite{ref_virchow2}. These foundation models have demonstrated exceptional generalization capabilities and strong performance in diverse downstream histopathology tasks, such as cancer classification, cancer grading, and prognosis prediction. Nonetheless, most existing studies that utilise foundation models have predominantly focused on classification tasks involving aggregation of patch-level features from WSIs. To date, exploration of the use of foundation models for high-resolution pixel-level segmentation tasks remains limited. Recently, CellViT++ \cite{ref_cellvit++} has emerged as one of the first approaches to utilise foundation models for nuclei segmentation, achieving SOTA performance on multiple cell segmentation benchmarks. However, to our knowledge, currently there is no published study that directly investigates the effectiveness of pathology foundation models for tissue segmentation on a large scale.

\subsection{Main Contributions}
Our main contributions are summarised as follows:
\begin{itemize}
    \item We propose a novel deep learning network that integrates the Virchow2 pathology foundation model with an EfficientNetV2 \cite{ref_efficientnetv2_paper} based encoder-decoder architecture (Efficient-UNet) to segment five tissue classes in H\&E-stained images of melanoma.  
    \item We introduce a dual stage loss strategy that provides intermediate and final supervision, which encourages consistent feature learning throughout the network.
    \item We validate our method in the PUMA Challenge\cite{ref_puma_paper}, where it achieved first place in the tissue segmentation task.
\end{itemize}

\section{Method}

\subsection{PUMA Dataset}
The dataset used in this study is sourced from the PUMA Challenge \cite{ref_puma_paper}. The organisers collected a total of 310 H\&E-stained region-of-interest (ROI) images. These include 155 primary and 155 metastatic melanoma samples. All ROIs were scanned at $40\times$ magnification (0.23 $\mu m$ per pixel), with a spatial resolution of $1024 \times 1024$ pixels. Both tissue and nuclei annotations were initially created by a medical expert and subsequently reviewed and refined by a dermatopathologist. While all slides were digitised at a single melanoma referral centre, 76 cases were referral consultations from other hospitals, which may introduce variability in staining due to differing processing techniques.
\subsubsection{Training Dataset}
The released training set includes 103 primary and 102 metastatic melanoma ROIs (one image was removed because of missing annotations). We use this dataset to train our models using 5-fold cross-validation.
\subsubsection{Test Dataset}
Two separate test sets are used for evaluation: a preliminary test set consists of 10 ROIs: 5 primary and 5 metastatic melanomas for initial validation. A final test set consists of 94 ROIs: 47 primary and 47 metastatic ROIs for the final evaluation to determine the challenge winners.
\subsubsection{Tissue Classes} The annotations include six tissue classes: tumour, stroma, epidermis, necrosis, blood vessels and background. Among these, tumour and stroma occupy the majority of tissue area, with epidermis, necrosis, and blood vessels being the minority classes. \newline
In this paper, we report the performance of our proposed method, alongside our baseline models and top-performing models from other teams, on the preliminary and final test sets. Background class is excluded from evaluation.

\subsection{Overview of the Proposed Framework}  
\begin{figure}[h!t]
\includegraphics[width=\textwidth]{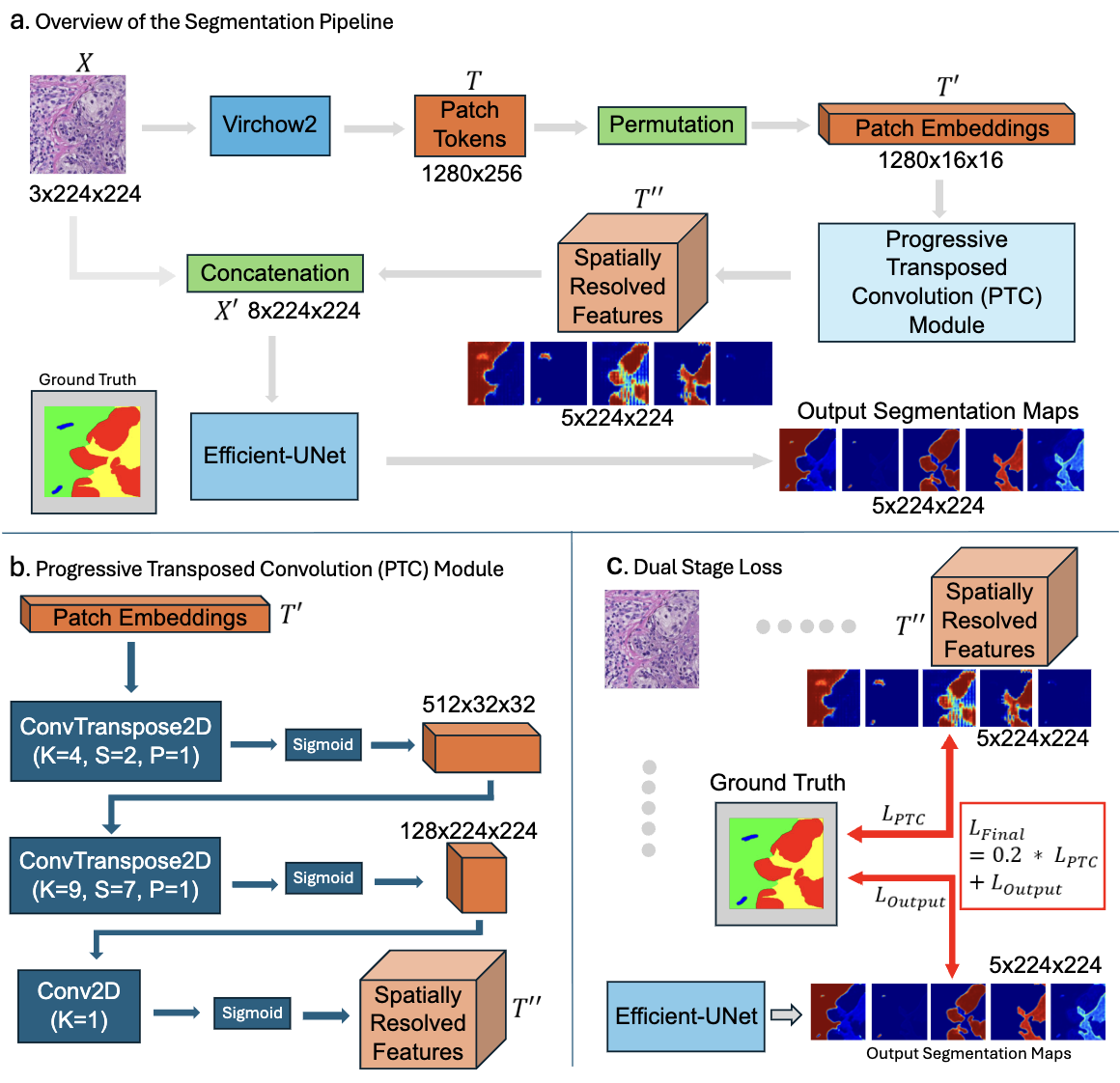}
\caption{Illustrations of our proposed method. \textbf{a.} The end-to-end segmentation pipeline: the input RGB image is passed through Virchow2 to extract patch embeddings. These embeddings are processed by the Progressive Transposed Convolution (PTC) module to generate spatially resolved feature maps. The resulting features are then concatenated with the original RGB image and passed to an Efficient-UNet for final segmentation. \textbf{b.} Architecture of the PTC module: patch embeddings are upsampled and reduced in dimensionality through two consecutive ConvTranspose2D layers, followed by a Conv2D layer to produce the final spatially resolved output with the desired number of channels. \textbf{c.} Illustration of the dual stage loss: losses are computed at both the intermediate stage after the PTC module ($L_{PTC}$) and the final output stage ($L_{Output}$). This encourages consistent feature learning throughout the network.} 
\label{network}
\end{figure}

Our goal is to design a neural network capable of learning robust and generalisable features for accurate tissue segmentation in melanoma. To achieve this, we use Virchow2 \cite{ref_virchow2}, a 632-million-parameter ViT trained on 3.1 million histopathology WSIs, as a frozen feature extractor. Patch tokens from the pre-trained Virchow2 provide guidance for segmentation, which we integrate into an EfficientNetV2 based encoder-decoder segmentation model (Efficient-UNet). Another motivation for selecting Virchow2 is its ability to produce patch tokens out-of-the-box.
\begin{enumerate}
    \item \textbf{Input Image}: Given an input RGB image of shape $3\times1024\times1024$ ($C \times H \times W$) pixels at $40\times$ magnification, it is downsampled to $3\times224\times224$ pixels using bilinear interpolation, we obtain $X$.
    \item \textbf{Patch Token Extraction}: $X$ is passed to Virchow2, which produces a sequence of patch tokens $T$ of dimension $1280\times256$.
    \item \textbf{Permutation}: We rearrange $T$ into a multidimensional grid $T'$ of shape $1280\times16\times16$ ($C \times H \times W$). This facilitates convolution operations.
    \item \textbf{Progressive Transposed Convolution (PTC) Module}: $T'$ is progressively upsampled by the PTC Module. The output is a $5\times224\times224$ ($C \times H \times W$) spatially resolved feature map, $T''$
    \item \textbf{Feature Concatenation}: $T''$ is concatenated with the original image patch $X$, producing $X'$, a tensor of dimension $8\times224\times224$ ($C \times H \times W$). This step preserves both token-based features from Virchow2 and raw context from the input image.
    \item \textbf{Efficient\-UNet Segmentation}: $X'$ is passed to Efficient\-UNet, an encoder-decoder network, with an EfficientNetV2-M backbone. It outputs a segmentation map of dimension $5\times224\times224$ ($C \times H \times W$), corresponding to five target tissue classes.
    \item \textbf{Dual Stage Loss}: Loss is computed at two stages, after the PTC Module ($L_{PTC}$) and at the final output stage ($L_{Output}$). $L_{PTC}$ and $L_{Output}$ are combined using a weighted sum. Details about the loss functions are explained in Section \ref{loss_section}.
\end{enumerate}
A visualisation of these steps is shown in Fig \ref{network}.

\subsection{Progressive Transposed Convolution (PTC) Module}
We propose a PTC module to restore spatial information from patch tokens. This module consists of three convolution layers with sigmoid activations, which upsample patch tokens from $16\times16$ to $224\times224$ while reducing the channel dimension from 1280 to 5. We choose five output channels here because it matches the number of target tissue types, it provides a direct feature mapping to the segmentation network and it facilitates intermediate loss calculation (this is explained in Section \ref{loss_section}). \newline 
As illustrated in Fig \ref{network}.b, the patch tokens extracted by Virchow2 from an input RGB image are passed through two ConvTranspose2D layers. A final $1\times1$ convolution is applied to generate the 5-channel output corresponding to the target tissue classes. We employ sigmoid activations following each convolution block; in our experiments, this significantly stabilizes training by constraining intermediate feature values within [0,1], which mitigates exploding gradients compared to ReLU activation. \newline
We also considered fusing the spatially resolved features with Efficient-UNet at its bottleneck layer, but the network showed poor performance in our internal cross-validation.

\subsection{Efficient-UNet}
We constructed an encoder–decoder network with U-Net–style skip connections \cite{ref_unet_paper}, replacing the encoder with an EfficientNetV2-M \cite{ref_efficientnetv2_paper} backbone pre-trained on ImageNet, and added Spatial and Channel Squeeze \& Excitation (SCSE) modules \cite{SCSE} into the decoder blocks. We refer to this segmentation model as Efficient-UNet. Similar approaches have been explored in other studies \cite{efficientunet_1,efficientunet_2,efficientunet_3}.

\subsection{Dual Stage Loss} \label{loss_section}
\subsubsection{Loss Function}
The loss function is a combination of Dice loss and Focal loss ($Dice_{FL}$), with a larger weighting on Dice loss. The weighting factor was chosen empirically based on internal cross-validation. The equation of $Dice_{FL}$ is shown in equation \ref{loss_eq}.
\begin{equation} \label{loss_eq}
    Dice_{FL} = 2 * Dice_{loss} + Focal_{loss}
\end{equation}
The equation of Dice loss ($Dice_{loss}$) is shown in equation \ref{dice_eq}. It is calculated as one minus the Dice coefficient between the ground truth $g$ and the predicted probability $p$ for each pixel $i$, we set $\epsilon=1e^{-6}$ to avoid division by zero.
\begin{equation} \label{dice_eq}
    Dice_{loss} = 1- Dice
                = 1 - \frac{2 \sum_i p_i g_i}{\sum_i p_i + \sum_i g_i + \epsilon}
\end{equation}
The equation of Focal loss ($Focal_{loss}$) is shown in equation \ref{focal_eq}. We set $\alpha=0.3$ and $\gamma=3.5$, based on our internal cross-validation experiments.
\begin{equation} \label{focal_eq}
Focal_{loss} = \begin{cases}
-\alpha (1-p_i)^\gamma \log{p_i} &\text{if $g_i$ = 1}\\
-(1-\alpha)p_i^\gamma \log{1-p_i} &\text{if $g_i$ = 0}
\end{cases}
\end{equation}

With a combination of Dice loss and Focal loss, we effectively optimise region-based overlap, while providing more emphasis on minority classes.

\subsubsection{Loss Computation}
We employ a dual stage loss strategy to reinforce robust feature learning throughout the pipeline, as illustrated in Fig\ref{network}.c. We introduce an intermediate loss after the PTC Module computed using $Dice_{FL}$ ($L_{PTC}$), which is combined with the output loss from the final segmentation layer computed using $Dice_{FL}$ ($L_{Output}$) through a weighted sum. \newline
Following the PTC Module, we obtain spatially resolved features of dimension $5\times224\times224$. Supervising this initial output against the ground truth provides an explicit training signal that encourages the PTC Module to learn task-relevant features. Additionally, gradients can flow directly back to the PTC Module without being diluted through the full segmentation network.
\begin{equation}
    L_{Final} = 0.2 * L_{PTC} + L_{Output}
\end{equation}
We combine $L_{PTC}$ and $L_{Output}$ using a weighted sum, where a weight of 0.2 on $L_{PTC}$ was selected empirically based on internal cross-validation. A larger weight on $L_{PTC}$ was found to destabilize training and make it difficult to converge, it is likely because the PTC Module alone cannot be expected to generate accurate segmentation maps purely from patch tokens. By keeping this weight small, we ensure the intermediate stage remains guided by useful feedback while the Efficient-UNet is ultimately responsible for making the final predictions.

% \subsection{Evaluation Metrics}
% The Dice Similarity Coefficient (Dice) is used to evaluate segmentation quality. The micro-average dice score is calculated by aggregating all tissue segmentations along one axis, then averaging the dice score across all tissue classes.

\subsection{Implementation Details}

\subsubsection{Hardware}: We trained our model on a high-performance compute node using two CPU cores (Intel(R) Xeon(R) Platinum 8168), 20 gigabytes (GBs) of RAM, and one Nvidia Tesla V100 GPU with 32 GBs of memory.

\subsubsection{Data Pre-Processing}: Each image and its corresponding mask are resized to $224\times224$ pixels by interpolation.

\subsubsection{Training}: We implemented all models using PyTorch version 2.5 and trained them using the AdamW optimizer with an inital learning rate of 0.001 and a weight decay of 0.005, the batch size was set to 24. Early stopping was used to prevent overfitting. On average the models take 30 minutes to finish training.

\subsubsection{Weighted Sampling}: We implemented a pixel-level weighted sampling strategy. Patches containing more tissue of rare classes are assigned higher sampling probabilities, this ensures that images with necrosis, blood vessels or epidermis are more frequently included during training. 

\subsubsection{Data Augmentation}: We apply a wide range of data augmentations during training, this includes random RGB shift, random hue saturation and value (HSV) shift, Gaussian blur and sharpening, image compression, random brightness and contrast adjustments, random shifts and scaling,  random 90-degree rotations and horizontal/vertical flips.

\subsubsection{Post-Processing}: We apply argmax on the output segmentation map to select the tissue class with the highest probability at each pixel. To remove small holes and spurious objects, we perform morphological opening followed by morphological closing using a circular kernel of size 13 pixels.

\section{Results}
We tested our proposed method on the PUMA Challenge dataset, we report performance on both the preliminary test set and the final test set. For comparison, we also include the results of the top performing segmentation models submitted by the top three teams in the challenge, Team LSM, Team rictoo, and Team Biototem. While their methods remain unpublished at the time of writing, their results serve as strong reference points for benchmarking performance across both evaluation phases.

\subsection{Internal Cross-Validation}
We conducted 5-fold cross validation on the training set to evaluate model performance and select the best configuration. Our method was compared against two baseline models: Efficient-UNet and a U-Net with a Swin Transformer encoder \cite{ref_swin_paper} (Swin-UNet).  \newline

Our method achieved the highest average Dice score (68.23\%). We found that incorporating the dual-stage loss strategy led to an improvement in segmentation performance, from 66.84\% to 68.23\%. However, we also observed a large standard deviation in Dice scores for the rarer tissue classes, notably epidermis, blood vessels, and especially necrosis. This could be attributed to the class imbalance within the dataset. A summary of our internal cross validation results is presented in Table \ref{table_cv}. \newline

\begin{figure}[t!]
\includegraphics[width=\textwidth]{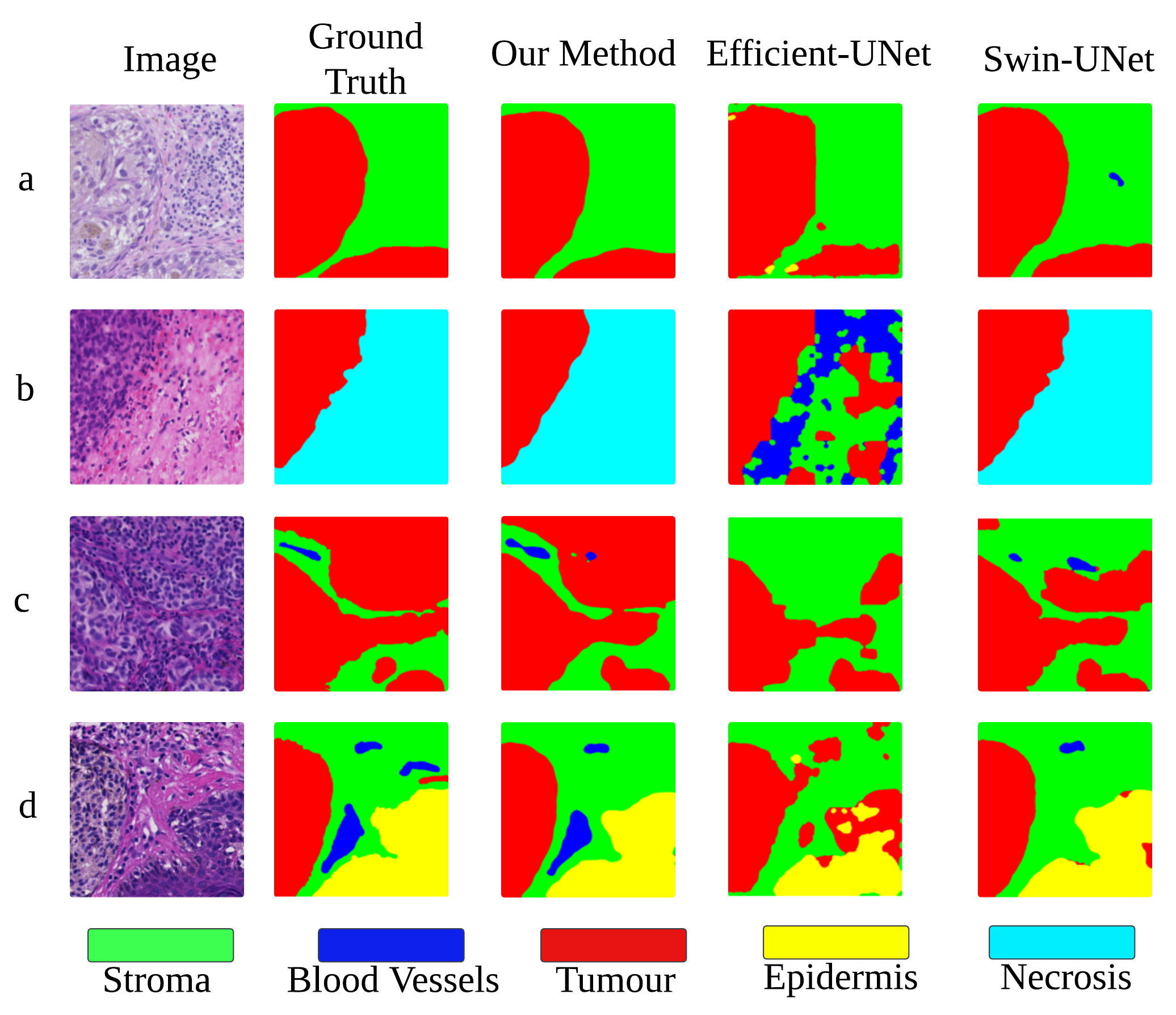}
\caption{Qualitative comparison of segmentation results between our proposed method, Efficient-UNet, and Swin-UNet on the internal 5-fold cross-validation set. \textbf{a}: Our method accurately segments tumour and stroma regions. Efficient-UNet introduces false positive epidermis, while Swin-UNet incorrectly predicts a blood vessel. \textbf{b}: Our method and Swin-UNet correctly segment tumour and necrosis regions. Efficient-UNet fails to detect necrosis and over-predicts multiple tissue types, especially blood vessels and stroma. \textbf{c}: Our method introduces a minor false positive blood vessel. Efficient-UNet and Swin-UNet misclassifies a large tumour region as stroma. \textbf{d}: Our method misses one blood vessel and a small tumour region. Efficient-UNet fails to detect blood vessels and confuses tumour with stroma and epidermis. Swin-UNet misses two blood vessels and partially misclassifies tissue boundaries.} \label{vis_comp}
\end{figure}

To further illustrate the performance differences between models, we present a visual comparison in Fig \ref{vis_comp}, showing segmentation outputs from our proposed method, Efficient-UNet, and Swin-UNet on image samples from the validation set. Overall, our method consistently produces more accurate segmentations, it particularly excels at tumour, stroma and epidermis regions. In contrast, the baseline models often misclassify tissue boundaries or completely fail to detect under-represented classes.

\begin{table}[h!t]
\caption{Summary of tissue segmentation results (Dice\%) from our \textbf{internal 5-fold cross validation}. *Our method trained using only $L_{output}$ loss rather than the proposed dual stage loss.}\label{table_cv}
\begin{tabularx}{\columnwidth}{|l|*{3}{>{\centering\arraybackslash}X}|}
\hline
Method &  Micro Average & Tumour& Stroma\\
\hline
\rowcolor{gray!30} 
\textbf{Our Method}& \textbf{68.23±6.61} & \textbf{93.73±1.44} & \textbf{83.60±5.46} \\
\hline
Our Method($L_{output}$ only)* & 66.84±8.05 & 93.09±1.42 & 82.72±3.19  \\
\hline
Efficient-UNet & 45.99±5.61 & 91.48±2.41 & 78.65±3.33  \\
\hline
Swin-UNet &  66.79±9.86 & 92.77±1.73 & 82.69±5.13  \\
\hline
\end{tabularx}
\newline
\vspace*{0.2 cm}
\newline
\begin{tabularx}{\columnwidth}{|l|*{3}{>{\centering\arraybackslash}X}|}
\hline
Method & Nercosis & Blood Vessels& Epidermis\\
\hline
\rowcolor{gray!30} 
\textbf{Our Method} & 33.86±32.85 & 52.51±6.84 & \textbf{77.49±3.78} \\
\hline
Our Method($L_{output}$ only)* & \textbf{39.11±31.09} & \textbf{55.24±6.02} & 72.48±8.51 \\
\hline
Efficient-UNet & 0.00±0.00 & 11.34±13.78 & 48.48±16.23 \\
\hline
Swin-UNet & 32.11±41.02 & 48.74±10.85 & 68.73±12.72 \\
\hline
\end{tabularx}
\end{table}

\subsection{Preliminary Test Phase}
In the preliminary test phase, we compared our method against Efficient-UNet, Swin-UNet, and nnUNet \cite{ref_nnunet_paper} which was trained by the PUMA Challenge organizers. Our approach consistently outperformed these architectures, achieving a micro Dice score of 73.41\%, with particularly strong performance in tumour, stroma, and necrosis segmentation. While our average score was slightly behind Team rictoo and Team LSM, this was primarily due to lower performance on the blood vessels class. However, it is important to note that the preliminary test set comprises only 10 images, therefore is not a representative indicator of large scale performance. A summary of these results is presented in Table \ref{table_preliminary}.

\begin{table}[h!t]
\caption{Summary of tissue segmentation results (Dice\%) on the PUMA Challenge \textbf{Preliminary Test Set}. *Methods have not been publicly released at the time of writing.}\label{table_preliminary}
\begin{tabularx}{\columnwidth}{|l|*{6}{>{\centering\arraybackslash}X}|}
\hline
Method &  Micro Average & Tumour& Stroma& Necrosis& Blood Vessels& Epidermis\\
\hline
\rowcolor{gray!30} 
Our Method&  73.41 & 93.36 & \textbf{86.41} & \textbf{64.97} & 33.02 & 89.29 \\
\hline
\textbf{Team rictoo}*& \textbf{75.83} & 91.47 & 83.80 & 59.52 & 59.13 & 85.23 \\
\hline
Team LSM*& 75.53 & 90.62 & 82.69 & 47.63 & \textbf{66.46} & 90.22 \\
\hline
Team Biototem*& 69.22 & 88.21 & 77.78 & 37.82 & 51.35 & 90.03 \\
\hline
Efficient-UNet &  59.01 & 91.53 & 84.93 & 0.0 & 26.27 & \textbf{92.32} \\
\hline
Swin-UNet &  69.21 & 87.97 & 78.46 & 52.03 & 36.97 & 90.64 \\
\hline
nnUnet (baseline) & 62.88 & \textbf{93.37} & 84.92 & 0.0 & 50.83 & 85.30 \\
\hline
\end{tabularx}
\end{table}

\begin{table}[h!t]
\caption{Summary of tissue segmentation results (Dice\%) on the PUMA Challenge \textbf{Final Test Set}. *Methods have not been publicly released at the time of writing. **Results reported by the authors using two significant figures.}\label{table_final}
\begin{tabularx}{\columnwidth}{|l|*{6}{>{\centering\arraybackslash}X}|}
\hline
Method &  Micro Average & Tumour& Stroma& Necrosis& Blood Vessels& Epidermis\\
\hline
\rowcolor{gray!30} 
\textbf{Our Method}& \textbf{78.23} & \textbf{93.58} & \textbf{83.59} & \textbf{82.04} & 45.70 & 86.26 \\
\hline
Team rictoo*& 63.26 & 91.77 & 81.49 & 15.19 & 47.22 & 80.63 \\
\hline
Team LSM*& 77.98 & 92.46 & 81.28 & 74.49 & \textbf{54.37} & \textbf{87.32} \\
\hline
Team Biototem*& 72.69 & 91.43 & 81.29 & 57.18 & 47.35 & 86.17 \\
\hline
nnUnet (baseline) & 55.48 & 91.09 & 78.51 & 1.52 & 34.99 & 71.30 \\
\hline
MaskFormer-UNI**&  44.00 & 86.00 & 62.00 & 9.00 & 1.00 & 64.00 \\
\hline
\end{tabularx}
\end{table}

\subsection{Final Test Phase}
Our method ranked first in the final test phase of the PUMA Challenge, with a micro-average Dice score of 78.23\%. It demonstrated strong segmentation performance across most tissue classes. This represents a substantial improvement over the nnUNet baseline (Dice=55.48\%) as well as team rictoo (Dice=63.26\%), both of which showed significant performance degradation on the final test set. Furthermore, the challenge organizers trained a MaskFormer \cite{ref_maskformer_paper} model using the UNI foundation model \cite{ref_UNI_paper} as backbone, but it performed poorly (Dice=44.00\%)\cite{ref_puma_paper}. This highlights the challenge of effectively integrating pathology foundation models into segmentation networks. A summary of the final challenge leaderboard is provided in Table \ref{table_final}.

%Remove the boundary

\subsection{Visualisation}
\begin{figure}[h!p]
\includegraphics[width=\textwidth]{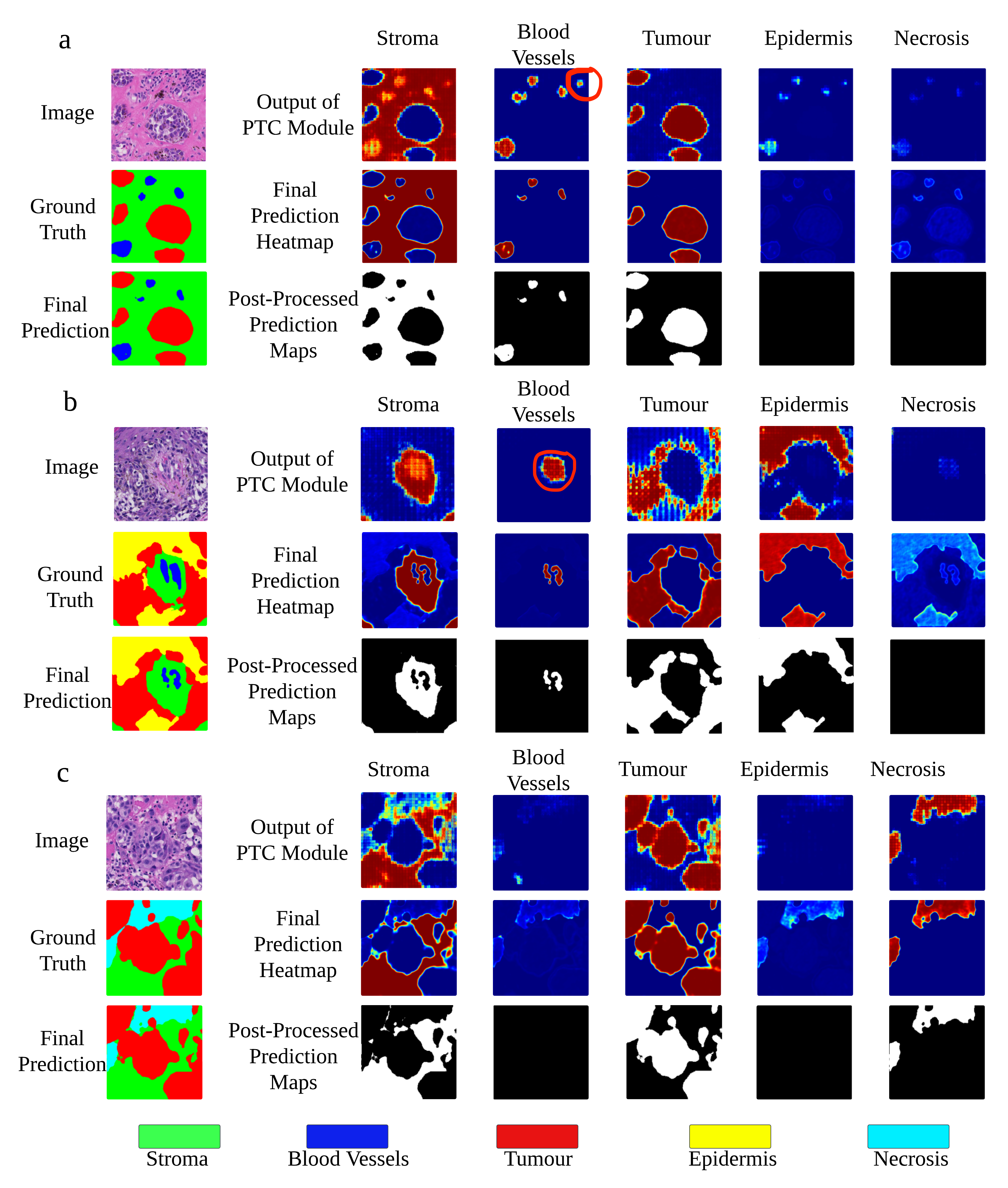}
\caption{\textbf{a, b, c}: Visualisation of example images predicted by our network. (\textbf{Top row}: Spatially resolved feature maps generated by the PTC Module from Virchow2 patch tokens, where each channel corresponds to one tissue class. \textbf{Middle row}: Probability maps produced by Efficient-UNet from the fusion of the spatially resolved feature maps and the RGB image. \textbf{Bottom row}: Final segmentation output after post-processing.) \textbf{a}: A false positive blood vessel instance (circled in red) can be seen in the output of the PTC Module, this is resolved by Efficient-UNet. \textbf{b}: Blood vessels prediction from the PTC Module falsely merged two instances into one (circled in red), this is resolved by Efficient-UNet. These show that the PTC output is used as a guidance rather than a hard constraint. \textbf{c}: Output from the PTC module is refined by Efficient-UNet.} \label{output_vis}
\end{figure}

To further validate the effectiveness of the patch tokens from Virchow2 and the ability of our Progressive Transposed Convolution (PTC) module to generate meaningful spatially resolved features, we visualise the PTC output as probability heat maps, where each of the five output channels corresponds to a specific tissue class: stroma, blood vessels, tumour, epidermis, and necrosis (see Fig \ref{output_vis}).

We observe that these spatially resolved feature maps closely resemble the ground truth annotations, which indicates the PTC module, when supervised with our dual-stage loss, is able to capture semantically relevant tissue representations at high resolution. 

However, as expected, these intermediate predictions are not always correct. For example, in Fig \ref{output_vis}.a, the spatially resolved features contain a false positive blood vessel instance in the top right corner (circled in red); Similarly, in Fig \ref{output_vis}.b, the spatially resolved features failed to separate instances of two blood vessels (circled in red). Since these features are fused with the original RGB image and passed through an Efficient-UNet, the network is able to use the PTC output as a guidance rather than as a hard constraint. This allows Efficient-UNet to correct errors, resulting in a final segmentation map that is both smooth and accurate, excluding the false positive blood vessel and separating instances of blood vessels in these examples.

\section{Inference Time}
We measured the inference time of our model and compared it against the baseline models Efficient-UNet and Swin-UNet. All models were evaluated on a workstation equipped with an NVIDIA RTX 3060 GPU. For each model, the forward pass was executed 100 times using the same image resolution (224x224), the average inference time per image along with the standard deviation are summarised in Table \ref{table_time}. As shown, our method incurs a higher inference cost compared to the baseline models, due to the overhead of incorporating Virchow2. Despite this, the inference time remains under 80 milliseconds per image, suggesting that the method is still practical for deployment.
\begin{table}[t!]
\caption{Average inference time per image (in seconds).}\label{table_time}
\begin{tabularx}{\columnwidth}{|l|*{1}{>{\centering\arraybackslash}X}|}
\hline
Method & Inference Time \\
\hline
\rowcolor{gray!30} 
Our Method& 0.0774 ± 0.0009 \\
\hline
Efficient-UNet & 0.0103 ± 0.0007 \\
\hline
Swin-UNet & 0.0113 ± 0.0005 \\
\hline
\end{tabularx}
\end{table}

\section{Conclusion}
In this work, we proposed a novel segmentation network that effectively harnesses the power of Virchow2, a pathology foundation model, for the segmentation of five tissue classes in H\&E-stained melanoma images. We introduce a Progressive Transposed Convolution (PTC) module to transform Virchow2 patch tokens into spatially resolved features. These features are then fused with the original RGB image and processed by an Efficient-UNet to produce precise segmentation maps. In addition, we employ a dual stage loss strategy that provides supervision at both the intermediate and the final stage of the network to encourage learning of consistent features throughout the pipeline. \newline

Quantitative results demonstrate that our model delivers strong performance and generalises effectively to unseen data. It achieved first place in the PUMA Challenge for the tissue segmentation task, significantly outperforming several baseline models. Furthermore, by examining the spatially resolved features, we observe that while Virchow2 is a powerful foundation model, it could still produce false predictions in challenging melanoma tissue regions. \newline

In future work, we aim to further validate our method on larger datasets such as the BCSS dataset \cite{bcss}, and conduct a detailed quantitative evaluation using only the spatially resolved features as the final output to better understand their predictive value. We also plan to perform an ablation study to assess the impact of different weighting configurations in the dual stage loss strategy, and to explore the use of alternative pathology foundation models, such as UNI \cite{ref_UNI_paper}, Phikon-v2 \cite{ref_phikonv2}, and TITAN \cite{ref_titan}. Additionally, we will investigate the effect of using different input image resolutions, as our current approach includes resizing input images from $1024\times1024$ to $224\times224$ which may distort smaller structures such as blood vessels. Using higher-resolution inputs or a multiscale approach may yield further performance gains. \newline

The original source code will be made publicly available on GitHub, the link can be found from the PUMA Challenge \cite{ref_puma_paper} website (\url{https://puma.grand-challenge.org/}). Moreover, after conducting detailed ablation studies as a part of our future work, we will integrate the full inference pipeline of an extended version of our model into TIAToolbox \cite{tiatoolbox} (\url{https://github.com/TissueImageAnalytics/tiatoolbox/tree/master/examples/inference-pipelines}), a cross-platform Python library offering easy-to-use APIs for whole-slide image analysis in computational pathology. These future developments aim to create a robust and extensible tissue segmentation framework to support large-scale studies of the tumour micro-environment.

\begin{credits}
\subsubsection{\ackname} JL is supported by the UK Engineering and Physical Sciences Research Council (EPSRC). YZ is funded by China Scholarship Council - University of Warwick Scholarship. CGCT is supported by the European Union Grant agreement ID: 101119427. SEAR reports financial support by the MRC (MR/X011585/1) and the BigPicture project,  which has received funding from the Innovative Medicines Initiative 2 Joint Undertaking under grant agreement No 945358.

\subsubsection{\discintname}
The authors have no competing interests.
\end{credits}
\bibliographystyle{splncs04}
\bibliography{main}

\begin{thebibliography}{10}
\providecommand{\url}[1]{\texttt{#1}}
\providecommand{\urlprefix}{URL }
\providecommand{\doi}[1]{https://doi.org/#1}

\bibitem{bcss}
Amgad, M., et~al., E.: Structured crowdsourcing enables convolutional segmentation of histology images. Bioinformatics  \textbf{35}(18),  3461--3467 (02 2019)

\bibitem{bruni2020tumor}
Bruni, D., Angell, H.K., Galon, J.: The immune contexture and immunoscore in cancer prognosis and therapeutic efficacy. Nature Reviews Cancer  \textbf{20},  662--680 (2020). \doi{10.1038/s41568-020-0285-7}

\bibitem{ref_UNI_paper}
Chen, R., Ding, T., Lu, M., Williamson, D., Jaume, G., Song, A., Chen, B., Zhang, A., Shao, D., Shaban, M., Williams, M., Oldenburg, L., Weishaupt, L., Wang, J., Vaidya, A., Le, L., Gerber, G., Sahai, S., Williams, W., Mahmood, F.: Towards a general-purpose foundation model for computational pathology. Nature Medicine  \textbf{30},  850--862 (03 2024). \doi{10.1038/s41591-024-02857-3}

\bibitem{ref_maskformer_paper}
Cheng, B., Misra, I., Schwing, A., Kirillov, A., Girdhar, R.: Masked-attention mask transformer for universal image segmentation (12 2021). \doi{10.48550/arXiv.2112.01527}

\bibitem{ref_titan}
Ding, T., Wagner, S.J., Song, A.H., Chen, R.J., Lu, M.Y., Zhang, A., Vaidya, A.J., Jaume, G., Shaban, M., Kim, A., Williamson, D.F.K., Chen, B., Almagro-Perez, C., Doucet, P., Sahai, S., Chen, C., Komura, D., Kawabe, A., Ishikawa, S., Gerber, G., Peng, T., Le, L.P., Mahmood, F.: Multimodal whole slide foundation model for pathology (2024), \url{https://arxiv.org/abs/2411.19666}

\bibitem{dosovitskiy2020image}
Dosovitskiy, A., Beyer, L., Kolesnikov, A., Weissenborn, D., Zhai, X., Unterthiner, T., Dehghani, M., Minderer, M., Heigold, G., Gelly, S., Uszkoreit, J., Houlsby, N.: An image is worth 16x16 words: Transformers for image recognition at scale. In: International Conference on Learning Representations (2021), \url{https://openreview.net/forum?id=YicbFdNTTy}

\bibitem{ref_phikonv2}
Filiot, A., Jacob, P., Kain, A.M., Saillard, C.: Phikon-v2, a large and public feature extractor for biomarker prediction (2024), \url{https://arxiv.org/abs/2409.09173}

\bibitem{ref_melanoma_staging}
Gershenwald, J., Scolyer, R., Hess, K., Sondak, V., Long, G., Ross, M., Lazar, A., Faries, M., Kirkwood, J., Mcarthur, G., Haydu, L., Eggermont, A., Flaherty, K., Balch, C., Thompson, J.: Melanoma staging: Evidence-based changes in the american joint committee on cancer eighth edition cancer staging manual: Melanoma staging: Ajcc 8 th edition. CA: A Cancer Journal for Clinicians  \textbf{67} (10 2017). \doi{10.3322/caac.21409}

\bibitem{efficientunet_2}
Gomroki, M., Hasanlou, M., Reinartz, P.: Stcd-effv2t unet: Semi transfer learning efficientnetv2 t-unet network for urban/land cover change detection using sentinel-2 satellite images. Remote Sensing  \textbf{15}(5) (2023). \doi{10.3390/rs15051232}, \url{https://www.mdpi.com/2072-4292/15/5/1232}

\bibitem{hendry2017assessing}
Hendry, S., et~al., S.: Assessing tumor-infiltrating lymphocytes in solid tumors: A practical review for pathologists and proposal for a standardized method from the international immunooncology biomarkers working group. Advances In Anatomic Pathology  \textbf{24}, ~1 (08 2017). \doi{10.1097/PAP.0000000000000162}

\bibitem{efficientunet_3}
Huo, G., Lin, D., Yuan, M.: Iris segmentation method based on improved unet++. Multimedia Tools and Applications  \textbf{81}(28),  41249--41269 (Nov 2022). \doi{10.1007/s11042-022-13198-z}, \url{https://doi.org/10.1007/s11042-022-13198-z}

\bibitem{ref_cellvit++}
Hörst, F., Rempe, M., Becker, H., Heine, L., Keyl, J., Kleesiek, J.: Cellvit++: Energy-efficient and adaptive cell segmentation and classification using foundation models (2025), \url{https://arxiv.org/abs/2501.05269}

\bibitem{ref_nnunet_paper}
Isensee, F., Jaeger, P., Kohl, S., Petersen, J., Maier-Hein, K.: nnu-net: a self-configuring method for deep learning-based biomedical image segmentation. Nature Methods  \textbf{18}, ~1--9 (02 2021). \doi{10.1038/s41592-020-01008-z}

\bibitem{ref_swin_paper}
Liu, Z., Lin, Y., Cao, Y., Hu, H., Wei, Y., Zhang, Z., Lin, S., Guo, B.: Swin transformer: Hierarchical vision transformer using shifted windows (03 2021). \doi{10.48550/arXiv.2103.14030}

\bibitem{efficientunet_1}
Pillai, M.B., Nair, J.J.: Nuclei segmentation using unet with efficientnetv2 as encoder. In: Tuba, M., Akashe, S., Joshi, A. (eds.) ICT Systems and Sustainability. pp. 603--613. Springer Nature Singapore, Singapore (2023)

\bibitem{tiatoolbox}
Pocock, J., Graham, S., Vu, Q.D., Jahanifar, M., Deshpande, S., Hadjigeorghiou, G., Shephard, A., Bashir, R.M.S., Bilal, M., Lu, W., Epstein, D., Minhas, F., Rajpoot, N.M., Raza, S.E.A.: {TIAToolbox as an end-to-end library for advanced tissue image analytics}. Communications Medicine  \textbf{2}(1), ~120 (sep 2022). \doi{10.1038/s43856-022-00186-5}, \url{https://www.nature.com/articles/s43856-022-00186-5}

\bibitem{ref_unet_paper}
Ronneberger, O., Fischer, P., Brox, T.: U-net: Convolutional networks for biomedical image segmentation (01 2015)

\bibitem{SCSE}
Roy, A.G., Navab, N., Wachinger, C.: Concurrent spatial and channel `squeeze {\&} excitation' in fully convolutional networks. In: Frangi, A.F., Schnabel, J.A., Davatzikos, C., Alberola-L{\'o}pez, C., Fichtinger, G. (eds.) Medical Image Computing and Computer Assisted Intervention -- MICCAI 2018. pp. 421--429. Springer International Publishing, Cham (2018)

\bibitem{saltz2018spatial}
Saltz, J., Gupta, R., et~al., L.H.: Spatial organization and molecular correlation of tumor-infiltrating lymphocytes using deep learning on pathology images. Cell Reports  \textbf{23}(1),  181--193.e7 (2018). \doi{https://doi.org/10.1016/j.celrep.2018.03.086}, \url{https://www.sciencedirect.com/science/article/pii/S2211124718304479}

\bibitem{ref_puma_paper}
Schuiveling, M., Liu, H., Eek, D., Breimer, G., Suijkerbuijk, K., Blokx, W., Veta, M.: A novel dataset for nuclei and tissue segmentation in melanoma with baseline nuclei segmentation and tissue segmentation benchmarks. GigaScience  \textbf{14} (01 2025). \doi{10.1093/gigascience/giaf011}

\bibitem{srinidhi2021deep}
Srinidhi, C.L., Ciga, O., Martel, A.L.: Deep neural network models for computational histopathology: A survey. Medical Image Analysis  \textbf{67},  101813 (2021). \doi{https://doi.org/10.1016/j.media.2020.101813}, \url{https://www.sciencedirect.com/science/article/pii/S1361841520301778}

\bibitem{ref_efficientnetv2_paper}
Tan, M., Le, Q.: Efficientnetv2: Smaller models and faster training (04 2021). \doi{10.48550/arXiv.2104.00298}

\bibitem{ref_taube}
Taube, J., Galon, J., Sholl, L., Rodig, S., Cottrell, T., Giraldo, N., Baras, A., Patel, S., Anders, R., Rimm, D., Cimino-Mathews, A.: Implications of the tumor immune microenvironment for staging and therapeutics. Modern Pathology  \textbf{31} (12 2017). \doi{10.1038/modpathol.2017.156}

\bibitem{ref_weiss}
Weiss, S., Han, S.W., Lui, K., Tchack, J., Shapiro, R., Berman, R., Zhong, J., Krogsgaard, M., Osman, I., Darvishian, F.: Immunologic heterogeneity of tumor infiltrating lymphocyte composition in primary melanoma. Human Pathology  \textbf{57} (07 2016). \doi{10.1016/j.humpath.2016.07.008}

\bibitem{ref_virchow2}
Zimmermann, E., Vorontsov, E., Viret, J., Casson, A., Zelechowski, M., Shaikovski, G., Tenenholtz, N., Hall, J., Fuchs, T., Fusi, N., Liu, S., Severson, K.: Virchow 2: Scaling self-supervised mixed magnification models in pathology (08 2024). \doi{10.48550/arXiv.2408.00738}

\end{thebibliography}

\end{document}